\documentclass[letterpaper, 12pt]{article}[2000/05/19]

\usepackage[english]{babel}

\usepackage{amsfonts,amsmath,amssymb,amsthm,latexsym,amscd,mathrsfs}
\usepackage{ifthen,cite}
\usepackage[bookmarksnumbered=true]{hyperref}
\hypersetup{pdfpagetransition={Split}}

\newcommand{\evenhead}{Author \ name}
\newcommand{\oddhead}{Article \ name}
\newcommand{\theArticleName}{Article name}

\newcommand{\FirstPageHeading}[1]{\thispagestyle{empty}%
\noindent\raisebox{0pt}[0pt][0pt]{\makebox[\textwidth]{\protect\footnotesize \sf }}\par}

\newcommand{\ArticleName}[1]{\renewcommand{\theArticleName}{#1}\vspace{-2mm}\par\noindent {\LARGE\bf  #1\par}}
\newcommand{\Author}[1]{\vspace{5mm}\par\noindent {\it #1} \par\vspace{2mm}\par}
\newcommand{\Address}[1]{\vspace{2mm}\par\noindent {\it #1} \par}
\newcommand{\Email}[1]{\ifthenelse{\equal{#1}{}}{}{\par\noindent {\rm E-mail: }{\it  #1} \par}}
\newcommand{\URLaddress}[1]{\ifthenelse{\equal{#1}{}}{}{\par\noindent {\rm URL: }{\tt  #1} \par}}
\newcommand{\EmailD}[1]{\ifthenelse{\equal{#1}{}}{}{\par\noindent {$\phantom{\dag}$~\rm E-mail: }{\it  #1} \par}}
\newcommand{\URLaddressD}[1]{\ifthenelse{\equal{#1}{}}{}{\par\noindent {$\phantom{\dag}$~\rm URL: }{\tt  #1} \par}}

\newcommand{\Keywords}[1]{\vspace{3mm}\par\noindent\hspace*{10mm}
\parbox{140mm}{\small {\it Key words:} \rm #1}\par}
\newcommand{\Classification}[1]{\vspace{3mm}\par\noindent\hspace*{10mm}
\parbox{140mm}{\small {\it 2000 Mathematics Subject Classification:} \rm #1}\vspace{3mm}\par}

\newcommand{\ShortArticleName}[1]{\renewcommand{\oddhead}{#1}}
\newcommand{\AuthorNameForHeading}[1]{\renewcommand{\evenhead}{#1}}

\long\def\@makecaption#1#2{
  \sbox\@tempboxa{\small \textbf{#1.}\ \ #2}%
  \ifdim \wd\@tempboxa >\hsize
    {\small \textbf{#1.}\ \ #2}\par \else
    \global \@minipagefalse
    \hb@xt@\hsize{\hfil\box\@tempboxa\hfil}%
  \fi \vskip\belowcaptionskip}

\def\numberwithin#1#2{\@ifundefined{c@#1}{\@nocounterr{#1}}{%
  \@ifundefined{c@#2}{\@nocnterr{#2}}{%
  \@addtoreset{#1}{#2}%
  \toks@\@xp\@xp\@xp{\csname the#1\endcsname}%
  \@xp\xdef\csname the#1\endcsname
    {\@xp\@nx\csname the#2\endcsname.\the\toks@}}}}
\def\E^#1{{\buildrel #1 \over\vee}}
\newtheorem{theorem}{Theorem}

\newtheorem{proposition}{Proposition}
{\theoremstyle{definition} 

\newtheorem{remark}{Remark}

}

\begin {document}

\FirstPageHeading{G. Borgioli and V. Gerasimenko}

\ShortArticleName{Quantum Dual BBGKY Hierarchy}

\ArticleName{On the Initial-Value Problem to the\\ Quantum Dual BBGKY Hierarchy}

\Author{G. Borgioli $^\dag$ and V. Gerasimenko $^\ddag$}

\AuthorNameForHeading{G. Borgioli and V. Gerasimenko}

\Address{$^\dag$Universit\`{a} degli Studi di Firenze,
         Via S.Marta, 3, 50139 Firenze, Italy}
\Address{$^\ddag$Institute of Mathematics of NAS of Ukraine,
         Tereshchenkivs'ka Str., 3,\\ 01601 Kyiv-4, Ukraine}
\EmailD{$^\dag$giovanni.borgioli@unifi.it, $^\ddag$gerasym@imath.kiev.ua}

{\vspace{6mm}\par\noindent\hspace*{8mm}
\parbox{140mm}{\small { $\quad$

We develop a rigorous formalism for the description
of the evolution of observables in quantum systems of particles. We
construct a  solution of the initial-value
problem to the quantum dual BBGKY hierarchy of equations as an
expansion over particle clusters whose evolution are governed by
the corresponding-order dual cumulant (dual semi-invariant) of the
evolution operators of finitely many particles. For
initial data from the space of sequences of bounded
operators the existence and uniqueness theorem is proved.


\Classification{35Q40; 47d06.}
 } }
}

\makeatletter
\renewcommand{\@evenhead}{
\hspace*{-3pt}\raisebox{-15pt}[\headheight][0pt]{\vbox{\hbox to \textwidth {\thepage \hfil \evenhead}\vskip4pt \hrule}}}
\renewcommand{\@oddhead}{
\hspace*{-3pt}\raisebox{-15pt}[\headheight][0pt]{\vbox{\hbox to \textwidth {\oddhead \hfil \thepage}\vskip4pt\hrule}}}
\renewcommand{\@evenfoot}{}
\renewcommand{\@oddfoot}{}
\makeatother
\newpage
\protect\tableofcontents
\newpage

\section{Introduction}

Evolution equations of quantum many-particle systems arise in many
problems of modern statistical mechanics.
\cite{CGP97, AL , Pe95 , AA}. In the theory of such equations, during the last decade, many new results have been obtained,
in particular concerning the fundamental problem of the
rigorous derivation of quantum kinetic equations \cite{Sp07}-\cite{FL}.

A description of quantum many-particle systems can be formulated in
terms of two sets of objects: observables and states. The mean value
defines a duality between observables and states and as a
consequence there exist two approaches to the description of the
evolution. Usually the evolution of many-particle systems is
described, in the framework of the evolution of states, by the BBGKY
hierarchy for marginal density operators. In the papers
\cite{Gol}-\cite{FL} a solution of the Cauchy problem to the quantum
BBGKY hierarchy is constructed in the form of iteration series for
initial data in the space of sequences of trace class operators. In
\cite{GerSh , GerS} for the quantum BBGKY hierarchy (for the
classical many-particle systems in \cite{GerR , GerRS}) a
solution is represented in the form of series over particle
clusters, whose evolution is described by the corresponding order
cumulant (semi-invariant) of evolution operators of finitely many
particles. Using an analog of Duhamel formulas, such a solution
expansion reduces to an iteration series, which is valid for a
particular class of initial data and interaction potentials. An
equivalent approach for the description of  many-particle system
evolution is given by the evolution of observables and by the dual
BBGKY hierarchy. For classical systems this approach is studied in
the paper \cite{BG}.

In this paper we consider the initial-value problem to the quantum
dual BBGKY hierarchy of equations describing the evolution of
observables for both finitely and infinitely many quantum particles,
obeying Maxwell-Boltzmann statistics. We construct a  solution of
the Cauchy problem in the form of an expansion over clusters of the
decreasing number of particles, whose evolution is governed by the
corresponding-order dual cumulant (dual semi-invariant) of the
evolution groups of operators of finitely many particles (groups of
operators for the Heisenberg equations).

We now outline the structure of the paper and the main results.

In Section 2 we define the dual BBGKY hierarchy for quantum systems
with a general type of an interaction potential between particles
and construct a solution of the Cauchy problem to this hierarchy. We
define the dual cumulants (the dual semi-invariants) of evolution
operators for the Heisenberg equations and investigate some of their
typical properties. The concept of cumulants of evolution operators
forms the basis for the solution expansions of various evolution
equations for infinitely many particles \cite{GerSh , GerRS}.
In Section 3 we study the
properties of a one-parameter group of automorphisms generated by a
solution of the initial-value problem to the dual BBGKY hierarchy in the space of sequences of bounded operators
and prove the existence and uniqueness theorem of a classical and a
generalized solution for the corresponding Cauchy problem. In
Section 4 we consider the description of evolution for
infinite-particle systems in the Heisenberg picture
and prove the existence of mean value functionals for observables. In the
final section we give some concluding remarks.

Let us introduce some preliminary definitions and results for
the description of quantum many-particle systems of a non-fixed number of particles.

\subsection{The initial-value problem to the Heisenberg equation}
We consider a quantum system of a non-fixed number of identical (spinless) particles
with unit mass $m=1$ in the space $\mathbb{R}^{\nu},$ $\nu\geq 1$
(in the terminology of statistical mechanics it is known as non-equilibrium grand canonical ensemble \cite{CGP97}).

Let a sequence $g=\big(I,g_{1},\ldots,$ $g_{n},\ldots \big)$ be an infinite sequence of self-adjoint bounded
operators $g_{n}$  defined on the Fock space
$\mathcal{F}_{\mathcal{H}}={\bigoplus\limits}_{n=0}^{\infty}\mathcal{H}^{\otimes n}$ over the Hilbert space
$\mathcal{H}$~ ($\mathcal{H}^{0}=\mathbb{C}$ and $I$ is a unit operator).
An operator $g_{n}$, defined in the $n$-particle
Hilbert space $\mathcal{H}_{n}=\mathcal{H}^{\otimes n}$, will be denoted by $g_{n}(1,\ldots,n)$.
For a system of identical particles obeying
Maxwell-Boltzmann statistics, one has $ g_{n}(1,\ldots,n)=g_{n}(i_1,\ldots,i_n)$ for any permutation
of indexes $\{i_{1},\ldots,i_{n}\}\in \{1,\ldots,n\}$.

Let the space $\mathfrak{L}(\mathcal{F}_\mathcal{H})$ be the space of sequences
$g=\big(I,g_{1},\ldots,$ $g_{n},\ldots \big)$ of
bounded operators $g_{n}$ ($I$ is a unit operator) defined on the Hilbert space
$\mathcal{H}_n$ and satisfying symmetry property:
$ g_{n}(1,\ldots,n)=g_{n}(i_1,\ldots,i_n)$, if $\{i_{1},\ldots,i_{n}\}\in \{1,\ldots,n\}$, with an operator norm \cite{Kato}.
We will also consider a more general space $\mathfrak{L}_{\gamma}(\mathcal{F}_\mathcal{H})$
with a norm
\begin{eqnarray*}
            \|g\|_{\mathfrak{L}_{\gamma} (\mathcal{F}_\mathcal{H})}:=
            \max\limits_{n\geq 0}~ \frac{\gamma^n}{n!}~\|g_{n}\|_{\mathfrak{L}(\mathcal{H}_{n})},
\end{eqnarray*}
where $0<\gamma<1$ and  $\| . \|_{\mathfrak{L}(\mathcal{H}_{n})}$ is an operator norm \cite{Kato}.

An observable of finitely many quantum particles is a sequence of self-adjoint
operators from $\mathfrak{L}_{\gamma}(\mathcal{F}_\mathcal{H})$.
The case of the unbounded observables can be reduced to the case under consideration \cite{DauL_5}.

The evolution of observables $A(t)=\big(I,A_{1}(t,1),\ldots,A_{n}(t,1,\ldots,n),\ldots \big)$ is described by the initial-value problem
to the Heisenberg equation \cite{DauL_5, Pe95}
\begin{eqnarray}
     \label{H-N1}  &&\frac{d}{d t}A(t)=\mathcal{N}A(t),\\
     \label{H-N12} &&A(t)|_{t=0}=A(0),
\end{eqnarray}
where, if $g\in \mathcal{D}(\mathcal{N})\subset \mathfrak{L}(\mathcal{F}_\mathcal{H})$,
the von Neumann operator $\mathcal{N}={\bigoplus\limits}_{n=0}^{\infty}
\mathcal{N}_{n}$ is defined by  the formula
\begin{eqnarray}\label{dkomyt}
   (\mathcal{N}g)_{n}:= -\frac{i}{\hbar}\big(g_{n}H_{n}-H_{n}g_{n}\big).
\end{eqnarray}
$h={2\pi\hbar}$ is the Planck constant
and the Hamiltonian $H={\bigoplus\limits}_{n=0}^{\infty}H_{n}$ in \eqref{dkomyt} is a self-adjoint operator with a domain of the definition
$\mathcal{D}(H)=\{\psi=\oplus_{n=0}^{\infty} \psi_{n}\in{\mathcal{F}_{\mathcal{H}}}\mid \psi_{n}\in\mathcal{D}(H_n)\subset\mathcal{H}_{n},
~{\sum\limits}_{n}\|H_{n}\psi_{n}\|^{2}<\infty\}\subset{\mathcal{F}_{\mathcal{H}}}$ \cite{Kato}.

Assume
$\mathcal{H}=L^{2}(\mathbb{R}^\nu)$, then an element
$\psi\in\mathcal{F}_{\mathcal{H}}={\bigoplus\limits}_{n=0}^{\infty}L^{2}(\mathbb{R}^{\nu n})$ is a sequence of functions
$\psi=\big(\psi_0,\psi_{1}(q_1),\ldots,\psi_{n}(q_1,\ldots,q_{n}),\ldots\big)$
such that $\|\psi\|^{2}=|\psi_0|^{2} + \sum_{n=1}^{\infty}\int dq_1\ldots dq_{n}$ $|\psi_{n}(q_1,\ldots,q_{n})|^{2}<+\infty.$
On the subspace $L^{2}_0(\mathbb{R}^{\nu n})\subset L^{2}(\mathbb{R}^{\nu n})$ of infinitely
differentiable functions with compact support the
$n$-particle Hamiltonian $H_{n}$ acts according to the formula ($H_{0}=0$)
\begin{equation}\label{H_Zag}
                H_{n}\psi_n = -\frac{\hbar^{2}}{2}
               \sum\limits_{i=1}^{n}\Delta_{q_i}\psi_n
               +\sum\limits_{k=1}^{n}\sum\limits_{i_{1}<\ldots<i_{k}=1}^{n}\Phi^{(k)}(q_{i_{1}},\ldots,q_{i_{k}})\psi_{n},
\end{equation}
where  $\Phi^{(k)}$ is a $k$-body interaction potential satisfying Kato conditions \cite{Kato}.

On the space $\mathfrak{L}(\mathcal{F}_\mathcal{H})$ a solution of the Cauchy problem to the Heisenberg
equation \eqref{H-N1} is determined by the following one-parameter family of operators
\begin{eqnarray}\label{grG}
\mathbb{R}^1\ni t\mapsto\mathcal{G}(t)g:= \mathcal{U}(t)g\mathcal{U}^{-1}(t),
\end{eqnarray}
where $\,\mathcal{U}(t)={\bigoplus\limits}_{n=0}^{\infty}\mathcal{U}_{n}(t)$  and operators
 $\mathcal{U}_{n}(t),\,\mathcal{U}_{n}^{-1}(t)$ are defined as follows
\begin{eqnarray}\label{evol_oper}
                &&\mathcal{U}_{n}(t):= e^{{\frac{i}{\hbar}}tH_{n}},\nonumber\\
                &&\mathcal{U}_{n}^{-1}(t):= e^{-{\frac{i}{\hbar}}tH_{n}}
\end{eqnarray}
and $\mathcal{U}_{0}(t)=I$ is a unit operator.
\begin{proposition}
On the space $\mathfrak{L}_{\gamma}(\mathcal{F}_\mathcal{H})$ one-parameter mapping \eqref{grG}
\begin{eqnarray*}
\mathbb{R}^1\ni t\mapsto\mathcal{G}(t)g:= \mathcal{U}(t)g\mathcal{U}^{-1}(t)
\end{eqnarray*}
defines an isometric $\ast$-weak continuous group of operators, i.e. it is a $C_{0}^{\ast}$-group.
The infinitesimal generator  $\mathcal{N}={\bigoplus\limits}_{n=0}^{\infty}~
\mathcal{N}_{n}$ of this group of operators is a closed operator for the $\ast$-weak topology
and on its domain of the definition $\mathcal{D}(\mathcal{N})\subset\mathfrak{L}_{\gamma}(\mathcal{F}_\mathcal{H})$
which everywhere dense for the $\ast$-weak topology, $\mathcal{N}$  is
defined as follows in the sense of the $\ast$-weak convergence of the space
$\mathfrak{L}_{\gamma}(\mathcal{F}_\mathcal{H})$
\begin{equation}\label{infOper1}
 \mathrm{w^{\ast}-}\lim\limits_{t\rightarrow 0}\frac{1}{t}\big(\mathcal{G}(t)g-g\big)=-\frac{i}{\hbar}(gH-Hg),
\end{equation}
where  $H=\bigoplus^{\infty}_{n=0}H_{n}$ is the Hamiltonian \eqref{H_Zag}
and the operator: $\mathcal{N}g=(-i/\hbar)(gH-Hg)$ is defined on the domain $\mathcal{D}(H)\subset\mathcal{F}_\mathcal{H}.$
\end{proposition}
The validity of this statement follows from properties of one-parameter
groups \eqref{evol_oper} \cite{DauL_5}, \cite{BR}.
Group of operators \eqref{grG} preserves the self-adjointness of operators.

On the space $\mathfrak{L}_{\gamma}(\mathcal{F}_\mathcal{H})$,  for an abstract initial-value problem  \eqref{H-N1}-\eqref{H-N12},
the following statement holds
\begin{proposition}
A solution of the initial-value problem
to the Heisenberg equation \eqref{H-N1}-\eqref{H-N12} is determined by
\begin{equation}\label{sH}
    A(t) = \mathcal{G}(t)A(0),
\end{equation}
where the one-parameter
family $\{\mathcal{G}(t)\}_{t\in \mathbb{R}}$ of operators is defined by expression \eqref{grG}.
For $A(0)\in\mathcal{D}(\mathcal{N})\subset\mathfrak{L}_{\gamma}(\mathcal{F}_\mathcal{H}),\; A(t)$
is a classical solution and for arbitrary initial data $A(0)\in\mathfrak{L}_{\gamma}(\mathcal{F}_\mathcal{H}),\;
A(t)$ is a generalized solution.
\end{proposition}

According general theorems about properties of the dual semigroup \cite{BR}, \cite{Pazy}, \cite{Gold}
the validity of these propositions follows also from the fact that the group $\{\mathcal{G}(t)\}_{t\in\mathbb{R}}$
is dual to the strong continuous group of the von Neumann equation defined on the space of sequences
of trace class operators \cite{DauL_5}.

We remark that, in the framework of kernels and symbols of operators, infinitesimal generator \eqref{infOper1}
and groups \eqref{evol_oper} are studied in \cite{BerSh}.

\subsection{Quantum many-particle systems}
A positive continuous linear functional on the space of observables $A(t)\in\mathfrak{L}(\mathcal{F}_\mathcal{H})$, whose
value is interpreted as its mean value (average value of an observable), is defined as follows \cite{BR}
\begin{equation}\label{averageD}
        \big\langle A(t)\big|D(0)\big\rangle:=\big(\sum\limits_{n=0}^{\infty}\frac{1}{n!}
        \mathrm{Tr}_{\mathrm{1,\ldots,n}}~D_{n}(0)\big)^{-1}
        \sum\limits_{n=0}^{\infty}\frac{1}{n!}
        \mathrm{Tr}_{\mathrm{1,\ldots,n}}~A_{n}(t,1,\ldots,n)D_{n}(0,1,\ldots,n),
\end{equation}
where $\mathrm{Tr}_{\mathrm{1,\ldots,n}}$ is the partial trace over $1,\ldots,n$ particles and ${\sum\limits}_{n=0}^{\infty}\frac{1}{n!}
        \mathrm{Tr}_{\mathrm{1,\ldots,n}}~D_{n}(0)$ is a normalizing
factor (grand canonical partition function). The sequences $D(0)=\big(I,D_{1}(0,1),$ $\ldots,D_{n}(0,1,\ldots,n),\ldots\big)$
of positive self-adjoint operators $D_{n}$, $n\geq 1$, (the density operators whose kernels are known
as the density matrices \cite{BerSh}) defined on the $n$-particle
Hilbert space $\mathcal{H}_{n}=\mathcal{H}^{\otimes n}=L^{2}(\mathbb{R}^{\nu n})$,
describe the states of a quantum system of non-fixed number of particles.

Usually it is assumed that the states belong to the space
$\mathfrak{L}^{1}_{\alpha}(\mathcal{F}_\mathcal{H})= {\bigoplus\limits}_{n=0}^{\infty}
~ \alpha^{n} \mathfrak{L}^{1}(\mathcal{H}_{n})$ of sequences
$f=\big(I,f_{1},\ldots,f_{n},\ldots\big)$ of trace class operators
$f_{n}=f_{n}(1,\ldots,n)\in\mathfrak{L}^{1}(\mathcal{H}_{n})$, satisfying the Maxwell-Boltzmann
statistics symmetry condition, equipped with the trace norm
\begin{eqnarray*}
            \|f\|_{\mathfrak{L}^{1}_{\alpha} (\mathcal{F}_\mathcal{H})}:=
            \sum\limits_{n=0}^{\infty}~ \alpha^{n} \|f_{n}\|_{\mathfrak{L}^{1}(\mathcal{H}_{n})}:=
            \sum\limits_{n=0}^{\infty}~ \alpha^{n}~\mathrm{Tr}_{\mathrm{1,\ldots,n}}|f_{n}(1,\ldots,n)|,
\end{eqnarray*}
where $\alpha>1$ is a real number, $\mathrm{Tr}_{1,\ldots,n}$ is the partial trace over $1,\ldots,n$ particles.
We will denote by $\mathfrak{L}^{1}_{\alpha, 0}$ the everywhere dense set
in $\mathfrak{L}^{1}_{\alpha}(\mathcal{F}_\mathcal{H})$ of finite sequences
of degenerate operators (finite-rank operators) \cite{Kato} with infinitely differentiable kernels with compact support.
The space $\mathfrak{L}^{1}_{\alpha}(\mathcal{F}_\mathcal{H})$ contains sequences
of operators more general than those determining the states of systems.
We will also consider the space $\mathfrak{L}^{1}(\mathcal{F}_\mathcal{H})= {\bigoplus\limits}_{n=0}^{\infty}
\mathfrak{L}^{1}(\mathcal{H}_{n})$.

If $ A(t)\in\mathfrak{L}_{\gamma}(\mathcal{F}_\mathcal{H})$ and $D(0)\in\mathfrak{L}^{1}_{\alpha}(\mathcal{F}_\mathcal{H})$,
then functional \eqref{averageD} exists.

It is well known that there exists an equivalent approach to the description of the evolution
of quantum systems, which is given by the evolution of states.
In this case the evolution of all possible states $D(t)=\big(I,D_{1}(t,1),\ldots,D_{n}(t,1,\ldots,n),\ldots \big)$
is described by the initial-value problem to the von Neumann equation \cite{DauL_5, Pe95}
\begin{eqnarray}
     \label{F-N1}  &&\frac{d}{d t}D(t)=-\mathcal{N}D(t),\\
     \label{F-N12} &&D(t)|_{t=0}=D(0),
\end{eqnarray}
where if $f\in \mathfrak{L}^{1}_{0}(\mathcal{F}_\mathcal{H})\subset\mathcal{D}(\mathcal{N})\subset
\mathfrak{L}^{1}(\mathcal{F}_\mathcal{H})$
the von Neumann operator $(-\mathcal{N})$ is defined by
\begin{eqnarray}\label{dkom}
   (-\mathcal{N}f)_{n}:= -\frac{i}{\hbar}\big(H_{n}f_{n}-f_{n}H_{n}\big),
\end{eqnarray}
The von Neumann equation \eqref{F-N1} is dual to
the Heisenberg equation \eqref{H-N1} with respect to bilinear form \eqref{averageD}.

For initial data $D(0)$ from the space $\mathfrak{L}^{1}(\mathcal{F}_\mathcal{H})$ the solution
\begin{eqnarray*}
D(t)=\mathcal{G}(-t)D(0)
\end{eqnarray*}
of the Cauchy problem to the von Neumann equation \eqref{F-N1}-\eqref{F-N12}
is determined by the following one-parameter family of operators
on $\mathfrak{L}^{1}(\mathcal{F}_\mathcal{H})$
\begin{eqnarray}\label{groupG}
\mathbb{R}^{1}\ni t\mapsto\mathcal{G}(-t)f:= \mathcal{U}(-t)f\mathcal{U}^{-1}(-t),
\end{eqnarray}
where the operators $\mathcal{U}_{n}(-t),\,\mathcal{U}_{n}^{-1}(-t)$ are defined by \eqref{evol_oper}.

The properties of a one-parameter family $\{\mathcal{G}(-t)\}_{t\in\mathbb{R}}$ of operators \eqref{groupG} follow from
the properties of groups \eqref{evol_oper}.
\begin{proposition}
On the space $\mathfrak{L}^{1}(\mathcal{F}_\mathcal{H})$ one-parameter mapping \eqref{groupG}
\begin{eqnarray*}
\mathbb{R}^{1}\ni t\mapsto\mathcal{G}(-t)f:= \mathcal{U}(-t)f\mathcal{U}^{-1}(-t)
\end{eqnarray*}
defines an isometric strongly continuous group of operators, i.e.
it is a $C_{0}$-group, which preserves positivity and self-adjointness of operators.
If $f\in\mathfrak{L}_{0}^{1}(\mathcal{F}_\mathcal{H})\subset\mathcal{D}(-\mathcal{N})\subset\mathfrak{L}^{1}(\mathcal{F}_\mathcal{H})$
then the infinitesimal generator: $-\mathcal{N}=\bigoplus^{\infty}_{n=0}(-\mathcal{N}_{n})$ of this group of operators is determined
in the sense of the norm convergence in the space $\mathfrak{L}^{1}(\mathcal{F}_\mathcal{H})$ as follows
\begin{equation}\label{infOper}
 \lim\limits_{t\rightarrow 0}\frac{1}{t}\big(\mathcal{G}(-t)f-f\big)=-\frac{i}{\hbar}(Hf-fH)=-\mathcal{N}f,
\end{equation}
where $H=\bigoplus^{\infty}_{n=0}H_{n}$ is the Hamiltonian \eqref{H_Zag} and
the operator: $(-i/\hbar)(Hf-fH)$ is defined on the domain $\mathcal{D}(H)\subset\mathcal{F}_\mathcal{H}.$
\end{proposition}
Thus, if we consider group \eqref{grG} as the dual to group \eqref{groupG}, from the above Proposition it follows
the statement about properties of group \eqref{grG}.

For a system of a finite average number of particles there exists an equivalent
possibility to describe observables and states, namely, by sequences of
marginal observables (the so-called $s$-particle observables) $G(t)=\big(G_0,G_{1}(t,1),\ldots,G_{s}(t,1,\ldots,s),\ldots\big)$
and marginal states (or $s$-particle density operators) $F(0)=\big(I,F_{1}(0,1),\ldots,F_{s}(0,1,\ldots,n),\ldots\big)$
\cite{CGP97}, \cite{Pe95}. These sequences
are correspondingly introduced  instead of sequences $A(t)$ and $D(0)$, in such way that mean value \eqref{averageD}
does not change, i.e.
\begin{equation}\label{avmar}
        \big\langle A(t)\big|D(0)\big\rangle=\big\langle G(t)\big|F(0)\big\rangle=
        \sum\limits_{s=0}^{\infty}\frac{1}{s!}
        \mathrm{Tr}_{\mathrm{1,\ldots,s}}~G_{s}(t,1,\ldots,s)F_{s}(0,1,\ldots,s).
\end{equation}
Then marginal observables are defined by
\begin{eqnarray}\label{mo}
       G_{s}(t,Y):= \sum_{n=0}^s\,\frac{(-1)^n}{n!}\sum_{j_1\neq\ldots\neq j_{n}=1}^s
            A_{s-n}\big(t,Y\backslash \{j_1,\ldots,j_{n}\}\big),  \quad\!\! s\geq 1,
\end{eqnarray}
where $Y\equiv(1,\ldots,s)$ and the sequence $A(t)=\big(I,A_{1}(t,1),\ldots,A_{n}(t,1,\ldots,n),\ldots \big)$ is solution \eqref{sH}
of the initial-value problem
to the Heisenberg equation \eqref{H-N1}-\eqref{H-N12}, i.e. $A(t)=\mathcal{G}(t)A(0)$.

If we introduce the operator $\mathfrak{a^+}$ (an analog of the creation operator \cite{BG})
\begin{eqnarray}\label{oper_znuw}
         \big( \mathfrak{a}^+ g \big)_{s}(Y):=
\sum_{j=1}^s \, g_{s-1}(Y \backslash \{j \})
\end{eqnarray}
defined on $\mathfrak{L}_{\gamma}(\mathcal{F}_\mathcal{H})$, then expression \eqref{mo} can be rewritten in the following compact form
\begin{eqnarray*}
G(t)=e^{-\mathfrak{a}^+}\mathcal{G}(t)A(0).
\end{eqnarray*}

The evolution of marginal observables \eqref{mo} of both finitely and infinitely many quantum particles
is described by the initial-value problem to the dual BBGKY hierarchy.
For finitely many particles the quantum dual BBGKY hierarchy is equivalent
to the Heisenberg equation \eqref{H-N1} (the dual equation
to the Heisenberg equation \eqref{H-N1} is the von Neumann equation \eqref{F-N1}).
For systems of classical particles the dual BBGKY hierarchy was examined in \cite{CGP97}, \cite{BG}, \cite{GerR}.

\section{The quantum dual BBGKY hierarchy}
In order to approach the description of observables of quantum many-particle systems by the marginal observables
($s$-particle observables) we study the evolution of the system by means of the quantum dual BBGKY hierarchy.
We introduce such a hierarchy of evolution equations
and construct a solution of the Cauchy problem to this hierarchy.

\subsection{The initial-value problem to the quantum dual BBGKY \\hierarchy}
The evolution of marginal observables is described by the initial-value problem
for the following hierarchy of evolution equations
\begin{eqnarray}
\label{dh}
       &&\frac{d}{dt}G_{s}(t,Y)=\mathcal{N}_{s}(Y)G_{s}(t,Y)+\\
       &&\sum\limits_{n=1}^{s}\frac{1}{n!}
        \sum\limits_{k=n+1}^s \frac{1}{(k-n)!}\sum_{j_1\neq\ldots\neq j_{k}=1}^s
        \mathcal{N}_{\mathrm{int}}^{(k)}(j_1,\ldots,j_{k})
        G_{s-n}(t,Y\backslash\{j_1,\ldots,j_{n}\}),\nonumber
\end{eqnarray}
\begin{eqnarray}\label{dhi}
       G_{s}(t)\mid_{t=0}=G_{s}(0), \quad\!\quad\!\! s\geq 1,\quad\!\quad\!\!\quad\!\quad\!\!\quad\!\quad\!\!\quad\!\quad\!\!\quad\!\quad\!\!\quad\!\quad\!\!
\end{eqnarray}
where, on $\mathcal{D}(\mathcal{N}_k)\subset\mathfrak{L}(\mathcal{H}_k)$,
the operator $\mathcal{N}^{(n)}_{int}$ is defined by
\begin{eqnarray}\label{oper Nint2}
      \mathcal{N}^{(n)}_{\mathrm{int}}g_{n}:= -\frac{i}{\hbar}\big(g_{n}\Phi^{(n)}-\Phi^{(n)}g_{n}\big).
\end{eqnarray}
and the operator $\Phi^{(n)}$ is introduced in Hamiltonian \eqref{H_Zag}.
We refer to \eqref{dh} as the quantum dual BBGKY hierarchy since the canonical BBGKY hierarchy \cite{CGP97}
for marginal density operators is the dual hierarchy of evolution equations with respect to bilinear form \eqref{avmar}
to evolution equations \eqref{dh}.

In the case of two-body interaction potential \eqref{H_Zag}, hierarchy \eqref{dh} has the form
\begin{eqnarray}\label{dig}
        \frac{d}{dt}G_{s}(t,Y) = \mathcal{N}_{s}(Y)G_{s}(t,Y)+
         \sum_{j_1\neq j_{2}=1}^s
         \mathcal{N}_{\mathrm{int}}^{(2)}(j_1,j_{2})G_{s-1}(t,Y\backslash\{j_1\}), \quad s\geq 1,
\end{eqnarray}
where the operator $\mathcal{N}^{(2)}_{\mathrm{int}}$ is defined by \eqref{oper Nint2} for $n=2$.
For $\mathcal{H}=L^{2}(\mathbb{R}^\nu)$,  the evolution
of kernels of operators $G_{s}(t)$, $s\geq 1$, for equations \eqref{dig}, is given by
\begin{eqnarray*}
        &&i\hbar\frac{\partial}{\partial t}G_{s}(t,q_1,\ldots,q_s;q'_1,\ldots,q'_s)=
       \Big(-\frac{\hbar^2}{2}\sum\limits_{i=1}^s(-\Delta_{q_i}+\Delta_{q'_i})+\\
       &&\sum\limits_{1=i<j}^s\big(\Phi^{(2)}(q'_i-q'_j)-\Phi^{(2)}(q_i-q_j)\big)\Big)
        G_s(t,q_1,\ldots,q_s;q'_1,\ldots,q'_s)+\\
        &&\sum\limits_{1=i\neq j}^s\big(\Phi^{(2)}(q'_i-q'_j)-\Phi^{(2)}(q_i-q_j)\big)G_{s-1}(t,q_1,\ldots,\E^{j},\ldots,q_s;q'_1,\ldots,\E^{j},\ldots,q'_s),
\end{eqnarray*}
where $(q_1,\ldots,\E^{j},\ldots,q_s)\equiv(q_1,\ldots,q_{j-1},q_{j+1},\ldots,q_s).$
The dual BBGKY hierarchy for a system of classical particles stated in \cite{CGP97, BG}
is defined by similar recurrence evolution equations.

The quantum dual BBGKY hierarchy \eqref{dh} can be derived from the
sequence of the Heisenberg equations \eqref{H-N1} provided that observables of a system are described
in terms of  marginal operators ($s$-particle observables) \eqref{mo}.
\begin{remark}
Another way of looking to the derivation of the quantum dual BBGKY hierarchy consists
in the construction of adjoint (dual) equations
with respect to the bilinear form \eqref{avmar} to the quantum BBGKY hierarchy.
For the sequence $F(t)=\big(I,F_{1}(t,1),\ldots,$ $F_{s}(t,Y),\ldots\big)$ of $s$-particle density operators (marginal density operators)
$F_{s}(t,Y),$ $s\geq 1,$ the quantum BBGKY hierarchy
has the form \cite{GerSh}
\begin{eqnarray} \label{1b}
        &&\frac{d}{dt}F_{s}(t,Y)=-\mathcal{N}_{s}(Y)F_{s}(t,Y) +\nonumber\\
       &&\sum\limits_{k=1}^{s}\frac{1}{k!}\sum\limits_{i_1\neq\ldots\neq i_{k}=1}^{s}\,
       \sum\limits_{n=1}^{\infty}\frac{1}{n!}\mathrm{Tr}_{\mathrm{s+1,\ldots,s+n}}
       \big(-\mathcal{N}_{\mathrm{int}}^{(k+n)}\big)(i_1,\ldots,i_{k},X \backslash Y)
        F_{s+n}(t, X),
\end{eqnarray}
where $X \equiv\{1,\ldots,s+n\}$ and on $\mathfrak{L}_{0}^{1}(\mathcal{H}_{s+n})\subset\mathfrak{L}^{1}(\mathcal{H}_{s+n})$
the operator $\mathcal{N}^{(k+n)}_{\mathrm{int}}$ is defined by \eqref{oper Nint2}.
Indeed, hierarchy \eqref{dh} is dual to hierarchy \eqref{1b} with respect to bilinear form \eqref{avmar}.
\end{remark}
\begin{remark}
In the paper \cite{BG}, for classical systems of particles with a two-body interaction potential,
an equivalent representation for the dual hierarchy generator was used. In the case under consideration,
on the subspace $\mathcal{D}(\mathfrak{B^+})\subset\mathfrak{L}_{\gamma}(\mathcal{F}_\mathcal{H})$,
the generator has the following representation
\begin{eqnarray*}
{\mathfrak{B}}^{+}=\mathcal{N}+ \big[\mathcal{N}, \mathfrak{a}^{+} \big],
\end{eqnarray*}
where $\big[\, .\, , \,.\,\big]$ is a commutator and the operator $\mathfrak{a}^+$ is defined on the space
$\mathfrak{L}_{\gamma}(\mathcal{F}_\mathcal{H})$ by \eqref{oper_znuw}. In a general case
the generator of the quantum dual BBGKY hierarchy \eqref{dh} can be represented in the following form
\begin{eqnarray}\label{rdg}
{\mathfrak{B}}^{+}=\mathcal{N}+ \sum\limits_{n=1}^{\infty}\frac{1}{n!}
\big[\ldots\big[\mathcal{N},\underbrace{\mathfrak{a}^{+} \big],\ldots,\mathfrak{a}^{+}}_{\hbox{n-times}}\big]=
e^{-\mathfrak{a}^{+}}\mathcal{N}e^{\mathfrak{a}^{+}},
\end{eqnarray}
where the operators $e^{\pm \mathfrak{a}^{+}}$ are defined on the space $\mathfrak{L}_{\gamma}(\mathcal{F}_\mathcal{H})$
by the expansions
\begin{eqnarray*}
\big(e^{\pm \mathfrak{a}^{+}}g\big)_s(Y)=\sum_{n=0}^s\,\frac{(\pm 1)^n}{n!}\sum_{j_1\neq\ldots\neq j_{n}=1}^s
            g_{s-n}\big(Y\backslash \{j_1,\ldots,j_{n}\}\big),  \quad\!\! s\geq 1.
\end{eqnarray*}
Representation \eqref{rdg} is correct in consequence of definition \eqref{oper_znuw} of the operator $\mathfrak{a}^{+}$
and the validity of identity
\begin{eqnarray*}
&&\big(\big[\ldots\big[\mathcal{N},\underbrace{\mathfrak{a}^{+} \big],\ldots,\mathfrak{a}^{+}}_{\hbox{n-times}}\big]g\big)_s(Y)=\\
&&\sum\limits_{k=n+1}^s \frac{1}{(k-n)!}\sum\limits_{j_1\neq\ldots\neq j_{k}=1}^s
        \mathcal{N}_{\mathrm{int}}^{(k)}(j_1,\ldots,j_{k})g_{s-n}\big(Y\backslash\{j_1,\ldots,j_{n}\}\big),
\end{eqnarray*}
which for a two-body interaction potential reduces to the following one
\begin{eqnarray*}
\big(\big[\mathcal{N},\mathfrak{a}^{+}\big]g\big)_s(Y)=
\sum\limits_{j_1\neq j_{2}=1}^{s} \mathcal{N}_{\mathrm{int}}^{(2)}(j_1,j_{2})g_{s-1}\big(Y\backslash\{j_1\}\big).
\end{eqnarray*}
\end{remark}

\subsection{The formula for a solution and its representations}
We consider two different approaches to the construction of a solution of the quantum dual BBGKY hierarchy \eqref{dh}.
Since hierarchy \eqref{dh} has the structure of recurrence equations, we deduce that the solution
can be constructed by successive integration of the inhomogeneous Heisenberg equations.
Indeed, for solutions of the first two equations we obtain
\begin{eqnarray}
        &&G_{1}(t,1)=\mathcal{G}_{1}(t,1)G_{1}(0,1),\nonumber\\
        \label{iter2}
        &&G_{2}(t,1,2)=\mathcal{G}_{2}(t,1,2)G_{2}(0,1,2)+\\
        &&\int\limits_{0}^{t}dt_{1}\mathcal{G}_{2}(t-t_{1},1,2)
        \mathcal{N}^{(2)}_{\mathrm{int}}(1,2)\mathcal{G}_{1}(t_{1},1)\mathcal{G}_{1}(t_{1},2)\big(G_{1}(0,1)+ G_{1}(0,2)\big).\nonumber
\end{eqnarray}
Let us transform the second term on the right hand side of \eqref{iter2} for $G_{2}(t)$ as follows
\begin{eqnarray}\label{iter2kum}
        &&\int\limits_{0}^{t}dt_{1}\mathcal{G}_{2}(t-t_{1},1,2)
        \mathcal{N}^{(2)}_{\mathrm{int}}(1,2)\mathcal{G}_{1}(t_{1},1)\mathcal{G}_{1}(t_{1},2)\\
        &&=-\int\limits_{0}^{t}dt_{1}\frac{d}{dt_{1}}\big(\mathcal{G}_{2}(t-t_{1},1,2)
        \mathcal{G}_{1}(t_{1},1)\mathcal{G}_{1}(t_{1},2)\big)=
        \mathcal{G}_{2}(t,1,2)-\mathcal{G}_{1}(t,1)\mathcal{G}_{1}(t,2)\nonumber.
\end{eqnarray}
The operator $\mathcal{G}_{2}(t,1,2)-\mathcal{G}_{1}(t,1)\mathcal{G}_{1}(t,2):=  \mathfrak{A}_{2}^{+}(t,1,2)$
in equality \eqref{iter2kum} is the $2nd$-order dual cumulant of evolution operators $\mathcal{G}(t)=\bigoplus^{\infty}_{n=0}\mathcal{G}_n(t)$
\eqref{grG}. Equality \eqref{iter2kum}
is an analog of the Duhamel formula \cite{BanArl},
which holds rigorously, for example, for bounded interaction potentials.

Thus, for solutions of the first two equations of hierarchy \eqref{dh} we finally obtain
\begin{eqnarray*}
&&G_{1}(t,1)=\mathfrak{A}_{1}^{+}(t,1)G_{1}(0,1),\\
&&G_{2}(t,1,2)=\mathfrak{A}_{1}^{+}\big(t,(\{1,2\})_1\big)G_{2}(0,1,2)+ \mathfrak{A}_{2}^{+}(t,1,2)\big(G_{1}(0,1)+ G_{1}(0,2)\big),
\end{eqnarray*}
where we introduced the $1st$-order dual cumulant $\mathfrak{A}_{1}^{+}\big(t,(\{1,2\})_1\big):= \mathcal{G}_{2}(t,1,2)$
and the notation $(\{1,2\})_1$ denotes a set composed by only one of
elements 1 and 2 and it is similarly generalized in next formula \eqref{hsol}.

Making use of transformations similar to \eqref{iter2kum}, for $n>2$, a solution of equations \eqref{dh}, constructed by
successive integration of the inhomogeneous Heisenberg equations, is represented by the following expansions
\begin{eqnarray}\label{hsol}
       G _{s}(t,Y)=\sum_{n=0}^s\,\frac{1}{(s-n)!}\sum_{j_1\neq\ldots\neq j_{s-n}=1}^s
      \mathfrak{A}_{1+n}^+ \big(t,(Y\backslash X)_1,X\big) \,
      G_{s-n}(0,Y\backslash X),  \quad\!\! s\geq 1.
\end{eqnarray}
Here we used some abridged notations: $Y\equiv(1,\ldots,s)$, $X\equiv Y\backslash\{j_1,\ldots,j_{s-n}\}$,
the set $(Y\backslash X)_1$ consists of one element from $Y\backslash X=(j_1,\ldots,j_{s-n})$,
i.e. the set $(j_1,\ldots,j_{s-n})$ is a connected subset of the partition $\mathrm{P}$ ($|\mathrm{P}|=1,\; |\mathrm{P}|$
denotes the number of considered partitions).
The evolution operator $\mathfrak{A}^+_{1+n}\big(t,(Y\backslash X)_1, X\big)$ in \eqref{hsol} is the $(1+n)th$-order dual cumulant $\mathfrak{A}_{1+n}^{+}(t), \, n\geq 0,$ of groups of operators \eqref{grG}
\begin{equation} \label{rozv_rec}
    \mathfrak{A}^{+}_{1+n}\big(t,(Y\backslash X)_1, X\big):=
    \sum\limits_{\mathrm{P}:\,\{(Y\backslash X)_1, X\}={\bigcup}_i X_i}
    (-1)^{\mathrm{|P|}-1}({\mathrm{|P|}-1})!\prod_{X_i\subset \mathrm{P}}\mathcal{G}_{|X_i|}(t,X_i),
\end{equation}
where ${\sum}_\mathrm{P} $
is the sum over all possible partitions $\mathrm{P}$ of the set $\{(Y\backslash X)_1,j_1,\ldots,j_{s-n}\}$ into
$|\mathrm{P}|$ nonempty mutually disjoint subsets  $ X_i\subset \{(Y\backslash X)_1, X\}$.
We consider some properties and examples of dual cumulants \eqref{rozv_rec}  in the next subsection.

Using the identity
\begin{eqnarray}\label{id}
\sum\limits_{n=0}^{s}\frac{1}{(s-n)!}
        \sum_{j_1\neq\ldots\neq j_{s-n}=1}^s g_{s-n}\big(j_1,\ldots,j_{s-n}\big)=
\sum\limits_{n=0}^{s}\frac{1}{n!}
        \sum_{j_1\neq\ldots\neq j_{n}=1}^s g_{s-n}\big(Y\backslash \{j_1,\ldots,j_{n}\}\big)
\end{eqnarray}
which is valid for the Maxwell-Boltzmann statistics symmetry property, expansion \eqref{hsol} can be rewritten
in the following an equivalent form
\begin{eqnarray*}
       &&G _{s}(t,Y)=\\
       &&\sum_{n=0}^s\,\frac{1}{n!}\sum_{j_1\neq\ldots\neq j_{n}=1}^s
      \mathfrak{A}_{1+n}^{+} \big(t,\big(Y \backslash \{j_1,\ldots,j_{n}\}\big)_1, j_1,\ldots,j_{n}\big) \,
      G_{s-n}(0,Y \backslash \{j_1,\ldots,j_{n}\}),  \, s\geq 1.
\end{eqnarray*}

The formula for a solution of the quantum dual BBGKY hierarchy \eqref{dh} can be also
derived from solution \eqref{sH} of the initial-value problem
to the Heisenberg equation \eqref{H-N1}-\eqref{H-N12} on the basis of expansions \eqref{mo}.

For sake of simplicity we consider the additive-type observables in capacity of initial data,
i.e. one-component sequences $G^{(1)}(0)=\big(0,a_{1}(1),0,\ldots\big)$.
In this case the
expansion for solution \eqref{hsol} attains the form
\begin{eqnarray}\label{af}
       G^{(1)}_{s}(t,Y)= \mathfrak{A}_{s}^{+}(t,1,\ldots,s)
       \sum_{j=1}^s a_{1}(j), \quad s\geq 1.
\end{eqnarray}
To determine the unknown evolution operator $\mathfrak{A}_{s}^{+}(t)$ from expansion \eqref{af},
we equate expansion \eqref{af} for the operator
$G^{(1)}_{s}(t,Y)$ with its representation by formula \eqref{mo} for the additive-type
observables $A(0)=\big(0,a_{1}(1),\ldots,\sum\limits_{i=1}^{n} a_{1}(i),\ldots\big)$, i.e.
\begin{eqnarray}\label{rr}
       &&\mathfrak{A}_{s}^+(t,1,\ldots,s)
       \sum_{j=1}^s a_{1}(j)= \\
       &&\sum_{n=0}^s\,\frac{(-1)^n}{n!}\sum_{j_1\neq\ldots\neq j_{n}=1}^s
            \mathcal{G}_{s-n}\big(t,Y\backslash \{j_1,\ldots,j_{n}\}\big)\sum\limits_{i\in Y\backslash \{j_1,\ldots,j_{n}\}}a_{1}(i),
            \quad\quad\!\! s\geq 1.\nonumber
\end{eqnarray}
Solving recursion relations \eqref{rr} for arbitrary operators ${\sum}_{j=1}^s a_{1}(j)$ we find that the $sth$-order dual cumulant $\mathfrak{A}_{s}^+(t,1,\ldots,s)$
is defined by expansion \eqref{cumulant}, which is particular case of \eqref{rozv_rec}.
For example, the first two equations from recursion relations \eqref{rr} have the form
\begin{eqnarray*}
&&\mathfrak{A}_{1}^+(t,1)a_{1}(1)=\mathcal{G}_{1}(t,1)a_{1}(1),\\
&&\mathfrak{A}_{2}^{+}(t,1,2)\big(a_{1}(1)+a_{1}(2)\big)=
\mathcal{G}_{2}(t,1,2)\big(a_{1}(1)+a_{1}(2)\big)-\mathcal{G}_{1}(t,1)a_{1}(1)-\mathcal{G}_{1}(t,2)a_{1}(2)=\\
&&\big(\mathcal{G}_{2}(t,1,2)-\mathcal{G}_{1}(t,1)\mathcal{G}_{1}(t,2)\big)\big(a_{1}(1)+a_{1}(2)\big).
\end{eqnarray*}
Hence a solution of the quantum dual BBGKY hierarchy \eqref{dh} is defined by expansion \eqref{hsol} where the evolution operators
$\mathfrak{A}_{1+n}^{+}(t), \, n\geq 0,$ are cumulants \eqref{rozv_rec} of groups \eqref{grG}.
\begin{remark}
In the paper \cite{BG}, for classical systems of particles,
an equivalent representation for the dual hierarchy solution was used. For the case under consideration
the solution expansion has the following representation
\begin{eqnarray}\label{rds}
&&G(t)=\mathcal{G}(t)G(0)+ \sum\limits_{n=1}^{\infty}\frac{1}{n!}
\big[\ldots\big[\mathcal{G}(t),\underbrace{\mathfrak{a}^{+} \big],\ldots,\mathfrak{a}^{+}}_{\hbox{n-times}}\big]G(0)=\\
&&e^{-\mathfrak{a}^{+}}\mathcal{G}(t)e^{\mathfrak{a}^{+}}G(0),\nonumber
\end{eqnarray}
where $\big[\, .\, , \,.\,\big]$ is a commutator, the operator $\mathfrak{a}^+$ is defined on the space
$\mathfrak{L}_{\gamma}(\mathcal{F}_\mathcal{H})$ by \eqref{oper_znuw} and
the one-parameter
family $\{\mathcal{G}(t)\}_{t\in \mathbb{R}}$ of operators is defined by expression \eqref{grG}.

Representation \eqref{rds} is true in consequence of definition \eqref{oper_znuw} of the operator $\mathfrak{a}^{+}$
and the validity of identity \eqref{id} and of the following
\begin{eqnarray}\label{rrr}
&&\big(\frac{1}{n!}\big[\ldots\big[\mathcal{G}(t),\underbrace{\mathfrak{a}^{+} \big],\ldots,\mathfrak{a}^{+}}_{\hbox{n-times}}\big]g\big)_s(Y)=\\
&&\frac{1}{(s-n)!}\sum_{j_1\neq\ldots\neq j_{s-n}=1}^s
      \mathfrak{A}_{1+n}^{+} \big(t,(Y\backslash X)_1,X\big) \,
      g_{s-n}(Y\backslash X),  \quad\!\! s\geq 1,\nonumber
\end{eqnarray}
where $X\equiv Y\backslash\{j_1,\ldots,j_{s-n}\}$ and $Y\backslash X=(j_1,\ldots,j_{s-n})$.
For example, if $n=1$ we have
\begin{eqnarray*}
&&\big(\big[\mathcal{G}(t), \mathfrak{a}^{+} \big]g\big)_s(Y)=\sum_{j=1}^s\big(\mathcal{G}_s(t,Y) -
\mathcal{G}_{s-1}(t,Y\backslash \{j\})\big)g_{s-1}(Y\backslash \{j\})= \\
&&\sum_{j=1}^s \mathfrak{A}_{2}^{+}\big(t,(Y\backslash \{j\})_1,j \big) \,
      g_{s-1}(Y\backslash \{j\})\, .
\end{eqnarray*}
Some details of the calculations for identity \eqref{rrr}
will be given in a general case in the Remark 4 of next subsection.
\end{remark}

\subsection{Dual cumulants of groups of evolution operators}
We now consider some properties of dual cumulants \eqref{rozv_rec}.
Let us  expand the groups
of operators $\mathcal{G}(t)=\bigoplus^{\infty}_{n=0}\mathcal{G}_n(t)$ \eqref{grG} over new
evolution operators as the following cluster expansions
\begin{eqnarray}\label{groupKlast}
    &\mathcal{G}_{n}(t,Y)=\sum\limits_{\mathrm{P}:Y ={\bigcup_i} X_i}\,
      \prod\limits_{X_i\subset \mathrm{P}}\mathfrak{A}_{|X_i|}^{+}(t,X_i),
      \quad\!\! n = |Y| \geq 0,\quad\!\!
\end{eqnarray}
where ${\sum}_\mathrm{P}$ is the sum over all possible partitions $\mathrm{P}$ of the set $ Y\equiv(1,\ldots,n) $
into $|\mathrm{P}|$ nonempty
mutually disjoint subsets $ X_i\subset Y.$
A solution of recurrence relations \eqref{groupKlast} is determined by the expansions \cite{GerR}
\begin{eqnarray}\label{cumulant}
        \mathfrak{A}_{n}^{+}(t,Y)
        =\sum\limits_{\mathrm{P}:Y ={\bigcup}_i X_i}(-1)^{|\mathrm{P}|-1}(|\mathrm{P}|-1)!
        \prod_{X_i\subset \mathrm{P}}\mathcal{G}_{|X_i|}(t,X_i),
        \quad\!\! n = |Y| \geq 0,
\end{eqnarray}
where the notations are similar to that in \eqref{groupKlast}.
We refer to the evolution operator $\mathfrak{A}_{n}^{+}(t)$   as the $nth$-order dual cumulant
(dual semi-invariant) of group of operators \eqref{grG}.

The simplest examples of dual cumulants \eqref{cumulant} have the form
\begin{eqnarray*}
    &&\mathfrak{A}^{+}_{1}(t,1)=\mathcal{G}_{1}(t,1),\\
    &&\mathfrak{A}^{+}_{2}(t,1,2)=\mathcal{G}_{2}(t,1,2)-\mathcal{G}_{1}(t,1)\mathcal{G}_{1}(t,2), \\
    &&\mathfrak{A}^{+}_{3}(t,1,2,3)=\mathcal{G}_{3}(t,1,2,3)-\mathcal{G}_{1}(t,3)\mathcal{G}_{2}(t,1,2)-
      \mathcal{G}_{1}(t,2)\mathcal{G}_{2}(t,1,3)\\
    &&\qquad\qquad\qquad-\mathcal{G}_{1}(t,1)\mathcal{G}_{2}(t,2,3)+2!\mathcal{G}_{1}(t,1)\mathcal{G}_{1}(t,2)\mathcal{G}_{1}(t,3).
\end{eqnarray*}

In the case of a quantum system of non-interacting particles  $(n\geq 2)$
we have: $\mathfrak{A}_{n}^{+}(t)=0.$
Indeed, for non-interacting quantum particles obeying Maxwell-Boltzmann statistics it holds
$$\mathcal{G}_{n}(t,1,\ldots,n)=\prod_{i=1}^{n}\mathcal{G}_{1}(t,i)$$
and hence
\begin{eqnarray*}
        \mathfrak{A}_{n}^{+}(t,Y)=
        \sum\limits_{\mathrm{P}:\,Y={\bigcup}_i X_i}
        (-1)^{| \mathrm{P}|-1}(|\mathrm{P}|-1)!\prod\limits_{X_{i}\subset \mathrm{P}}{\prod}_{l_i=1}^{|X_{i}|}\mathcal{G}_{1}(t,l_i)\\
      = \sum\limits_{k=1}^{n}(-1)^{k-1}\mathrm{s}(n,k)(k-1)!\prod_{i=1}^{n}\mathcal{G}_{1}(t,i)=0,
\end{eqnarray*}
where $\mathrm{s}(n,k)$ are the Stirling numbers of the second kind. Here the following equality is used
\begin{equation}\label{Stirl}
       \sum\limits_{\mathrm{P}:\,Y={\bigcup}_i X_i}
        (-1)^{| \mathrm{P}|-1}(|\mathrm{P}|-1)!=
        \sum\limits_{k=1}^{n}(-1)^{k-1}\mathrm{s}(n,k)(k-1)!=\delta_{n,1},
\end{equation}
where $\delta_{n,1}$ is a Kroneker symbol.

The generator of the $1st$-order dual cumulant is given by operator \eqref{dkomyt}, i.e.
\begin{equation*}
       \mathrm{w^{\ast}-}\lim\limits_{t\rightarrow 0}\frac{1}{t}\big(\mathfrak{A}_{1}^{+}(t,Y)-I\big) g_{n}(Y)
       =\mathcal{N}_{n}(Y)g_{n}(Y),
\end{equation*}
where for $g_n\in\mathcal{D}(\mathcal{N}_n)\subset\mathfrak{L}(\mathcal{H}_n)$ this limit exists
in the sense of the $\ast$-weak convergence of the space $\mathfrak{L}(\mathcal{H}_n)$ (Proposition 1).

In the general case an infinitesimal generator of the $nth$-order dual cumulant, $n\geq 2,$
is the operator $(\mathcal{N}^{(n)}_{\mathrm{int}})$
defined by $n$-body interaction potential \eqref{H_Zag}.
Indeed, according to equality \eqref{Stirl} in the sense of a point-by-point
convergence in the space $\mathfrak{L}(\mathcal{H}_{n})$
for the $nth$-order dual cumulant, $n\geq2,$ we have
\begin{eqnarray*}
      \lim\limits_{t\rightarrow 0} \frac{1}{t} \mathfrak{A}_{n}^{+}(t,Y) g_{n}(Y)
       = \sum\limits_{\mathrm{P}:\,Y={\bigcup}_i X_i}\hskip-2mm
        (-1)^{|\mathrm{P}|-1}(|\mathrm{P}| -1)!\hskip-1mm\sum\limits_{X_i\subset \mathrm{P}}
        (\mathcal{N}_{|X_i|}(X_i))g_{n}(Y)=\\
        \sum\limits_{\mathrm{P}:\,Y={\bigcup}_i X_i}
        (-1)^{|\mathrm{P}|-1}(|\mathrm{P}| -1)! \sum\limits_{X_i\subset \mathrm{P}}\,\,
        \sum\limits_{k=2}^{\mid X_i\mid}\,\,
        \sum\limits_{i_{1}<\ldots<i_{k}\in\{X_i\}}\big(\mathcal{N}^{(k)}_{\mathrm{int}}(i_1,\ldots,i_{k})\big)g_{n}(Y),
\end{eqnarray*}
where  $\mathcal{N}^{(n)}_{\mathrm{int}}$
is defined as in \eqref{oper Nint2}, i.e.
\begin{eqnarray*}
      \mathcal{N}^{(n)}_{\mathrm{int}}g_{n}:= -\frac{i}{\hbar}\big(g_{n}\Phi^{(n)}-\Phi^{(n)}g_{n}\big),
\end{eqnarray*}
where the operator $\Phi^{(n)}$ is defined in Hamiltonian \eqref{H_Zag}.

Summing the coefficients of every operator $\mathcal{N}^{(k)}_{\mathrm{int}}$ we deduce
\begin{eqnarray}\label{Nint}
        \mathrm{w^{\ast}-}\lim\limits_{t\rightarrow 0}\frac{1}{t}\mathfrak{A}_{n}^{+}(t,Y) g_{n}(Y)
       = \mathcal{N}^{(n)}_{\mathrm{int}}(Y)g_{n}(Y),
\end{eqnarray}

Thus, if $g_n\in\mathcal{D}(\mathcal{N}_n)\subset\mathfrak{L}(\mathcal{H}_n)$, the generator of the $nth$-order dual cumulant
is defined by \eqref{Nint}
in the sense of the $\ast$-weak convergence of the space $\mathfrak{L}(\mathcal{H}_n).$

We remark that, at the initial time  $t=0$, solution \eqref{hsol} satisfies
initial condition \eqref{dhi}. Indeed,
according to definitions \eqref{evol_oper} ($\mathcal{U}^{\pm 1}_{n}(0)=I$ is a unit operator) and of equality \eqref{Stirl}
for $n \geq 1$, we have
\begin{equation*}
    \mathfrak{A}^+_{1+n}\big(0,(Y\backslash X)_1, X\big)=
    \sum\limits_{\mathrm{P}:\,\{(Y\backslash X)_1, X\}={\bigcup}_i X_i}
    (-1)^{\mathrm{|P|}-1}({\mathrm{|P|}-1})!I=I \delta_{n,1}.
\end{equation*}

\begin{remark}
Cluster expansions \eqref{groupKlast} can be put at the basis
of all possible solution representations of the quantum dual BBGKY hierarchy \eqref{dh}. In fact
we can obtain representation \eqref{rds} solving
recurrence relations \eqref{groupKlast} with respect to the $1st$-order
dual cumulants for the separation terms, which are independent from the variable $Y\backslash X\equiv(j_1,\ldots,j_{s-n})$
\begin{eqnarray*}
    \mathfrak{A}_{1+n}^{+}\big(t,(Y\backslash X)_1, X \big)=
   \sum\limits_{\substack{Z\subset X}} \mathfrak{A}_{1}^{+}\big(t,Y\backslash X\cup Z \big)
    \sum\limits_{\mathrm{P}:\,X \backslash Z ={\bigcup\limits}_i X_i}
    (-1)^{|\mathrm{P}|}\,|\mathrm{P}| ! \,\, {\prod}_{i=1}^{|\mathrm{P}|} \mathfrak{A}_{1}^{+}\big(t,X_{i} \big),
\end{eqnarray*}
where ${\sum\limits}_{\substack{Z\subset X}}$ is a sum over all subsets $Z\subset X$ of the set $X$.
Then, taking into account the identity
\begin{eqnarray*}
\sum\limits_{\mathrm{P}:\,X \backslash Z ={\bigcup\limits}_i X_i}
    (-1)^{|\mathrm{P}|}\,|\mathrm{P}| ! \,\, {\prod}_{i=1}^{|\mathrm{P}|} \mathfrak{A}_{1}^{+}\big(t,X_{i} \big)g_{s-n}(Y\backslash X)=
\sum\limits_{\mathrm{P}:\,X \backslash Z ={\bigcup\limits}_i X_i}
    (-1)^{|\mathrm{P}|}\,|\mathrm{P}| ! \,g_{s-n}(Y\backslash X)
\end{eqnarray*}
and the equality
\begin{eqnarray}\label{e}
 \sum\limits_{\mathrm{P}:\,X \backslash Z =
 {\bigcup\limits}_i X_i}(-1)^{|\mathrm{P}|}\,|\mathrm{P}|!=(-1)^{|X \backslash Z|},
\end{eqnarray}
for expansion \eqref{hsol} we have
\begin{eqnarray*}
       G _{s}(t,Y)=\sum_{n=0}^s\,\frac{1}{(s-n)!}\sum_{j_1\neq\ldots\neq j_{s-n}=1}^s\,\,
      \sum\limits_{\substack{Z\subset X}}\,(-1)^{|X \backslash Z|}\, \mathcal{G}_{s-n+|Z|}\big(t,Y\backslash X\cup Z \big)
     \,G_{s-n}(0,Y\backslash X).
\end{eqnarray*}
Thus, as a result of the symmetry property for the Maxwell-Boltzmann statistics
and of definition \eqref{oper_znuw} of the operator $\mathfrak{a}^{+}$, we derive representation \eqref{rds}
\begin{eqnarray*}
       &&G(t)=\sum\limits_{n=0}^{\infty}\frac{1}{n!}\,\sum\limits_{k=0}^{n}\,(-1)^{n-k}\,\frac{n!}{k!(n-k)!}
       \,(\mathfrak{a}^{+})^{n-k}\mathcal{G}(t)(\mathfrak{a}^{+})^{k}G(0)=\\
       &&e^{-\mathfrak{a}^{+}}\mathcal{G}(t)e^{\mathfrak{a}^{+}}G(0).
\end{eqnarray*}

We can obtain one more representation for a solution of the initial-value problem to the quantum dual BBGKY hierarchy
if we express the dual cumulants $\mathfrak{A}_{1+n}^+(t), ~n\geq1,$ of group of operators \eqref{grG}
with respect to the $1st$-order and $2nd$-order dual cumulants. In fact it holds
\begin{eqnarray*}
    \mathfrak{A}_{1+n}^{+}\big(t,(Y\backslash X)_1, X \big)=
    \sum\limits_{\substack{Z\subset X,\\Z\neq \emptyset}} \mathfrak{A}_{2}^{+}\big(t,Y\backslash X,Z\big)
    \sum\limits_{\mathrm{P}:\,X \backslash Z ={\bigcup\limits}_i X_i}
    (-1)^{|\mathrm{P}|}\,|\mathrm{P}| ! \,\, {\prod}_{i=1}^{|\mathrm{P}|} \mathfrak{A}_{1}^{+}\big(t,X_{i} \big),
\end{eqnarray*}
where ${\sum\limits}_{\substack{Z\subset X,\\Z\neq \emptyset}}$ is a sum over all nonempty subsets $Z\subset X$ of the set $X$.
Then taking into account the identity
\begin{eqnarray*}
\sum\limits_{\mathrm{P}:\,X \backslash Z ={\bigcup\limits}_i X_i}
    (-1)^{|\mathrm{P}|}\,|\mathrm{P}| ! \,\, {\prod}_{i=1}^{|\mathrm{P}|}
    \mathfrak{A}_{1}^{+}\big(t,X_{i} \big)g_{s-n}\big(Y\backslash X\big)=
\sum\limits_{\mathrm{P}:\,X \backslash Z ={\bigcup\limits}_i X_i}
    (-1)^{|\mathrm{P}|}\,|\mathrm{P}| ! \,g_{s-n}(Y\backslash X)
\end{eqnarray*}
and equality \eqref{e}, we get the following representation for expansion \eqref{hsol}
\begin{eqnarray*}
    &&G _{s}(t,Y)=\mathfrak{A}_{1}^{+}\big(t,Y\big)G_{s}(0,Y)+\\
    &&\sum_{n=1}^s\,\frac{1}{(s-n)!}\,\sum_{j_1\neq\ldots\neq j_{s-n}=1}^s\,\,\,
      \sum\limits_{\substack{Z\subset X,\\Z\neq \emptyset}}\,(-1)^{|X\backslash Z|}\,\,
      \mathfrak{A}_{2}^{+}\big(t,Y\backslash X,Z\big) \,G_{s-n}(0,Y\backslash X),
\end{eqnarray*}
where $Y\equiv(1,\ldots,s)$, $X\equiv Y\backslash\{j_1,\ldots,j_{s-n}\}$, i.e. $Y\backslash X=(j_1,\ldots,j_{s-n})$.
\end{remark}

\section{A solution of the initial-value problem to the \\quantum dual BBGKY hierarchy}
One-parameter mapping generated by solution \eqref{hsol} is not strong continuous on the space
$\mathfrak{L}_{\gamma}(\mathcal{F}_\mathcal{H})$. Thus, the constructed solution satisfies the initial-value
problem to the dual BBGKY hierarchy \eqref{dh}-\eqref{dhi} only in the sense of the $\ast$-weak convergence of the space $\mathfrak{L}_{\gamma}(\mathcal{F}_\mathcal{H})$.

\subsection{A group of operators for the quantum dual BBGKY hierarchy}
On the space $\mathfrak{L}_{\gamma}(\mathcal{F}_\mathcal{H})$ solution \eqref{hsol} of the initial-value
problem to the dual BBGKY hierarchy \eqref{dh}-\eqref{dhi} is determined by a one-parameter mapping with the following properties.

\begin{theorem}
If $g\in\mathfrak{L}_{\gamma}(\mathcal{F}_\mathcal{H})$ and $\gamma<e^{-1}$, then the one-parameter mapping
\begin{eqnarray}\label{krut_rozv}
       &&\mathbb{R}^{1}\ni t\mapsto(U^{+}(t)g) _{s}(Y):= \\
      &&\sum_{n=0}^s\,\frac{1}{(s-n)!}\sum_{j_1\neq\ldots\neq j_{s-n}=1}^s
      \mathfrak{A}_{1+n}^+ \big(t,(Y\backslash X)_1,X\big) \,
      g_{s-n}(Y\backslash X),  \quad\!\! s\geq 1 \nonumber
\end{eqnarray}
is a $C_{0}^{\ast}$-group.
The infinitesimal generator  ${\mathfrak{B}}^{+}={\bigoplus\limits}_{n=0}^{\infty}
{\mathfrak{B}}^{+}_{n}$  of this group of operators is a closed operator for the $\ast$-weak topology and
on the domain of the definition $\mathcal{D}({\mathfrak{B}}^{+})\subset\mathfrak{L}_{\gamma}(\mathcal{F}_\mathcal{H})$
which is the everywhere dense set for the $\ast$-weak topology of the space $\mathfrak{L}_{\gamma}(\mathcal{F}_\mathcal{H})$
it is defined by the operator
\begin{eqnarray}\label{d}
       &&({\mathfrak{B}}^{+} g)_{s}(Y):= \mathcal{N}_{s}(Y)g_{s}(Y)+\\
       &&\sum\limits_{n=1}^{s}\frac{1}{n!}
        \sum\limits_{k=n+1}^s \frac{1}{(k-n)!}\sum_{j_1\neq\ldots\neq j_{k}=1}^s
        \mathcal{N}_{\mathrm{int}}^{(k)}(j_1,\ldots,j_{k})
        g_{s-n}(Y\backslash\{j_1,\ldots,j_{n}\}), \nonumber
\end{eqnarray}
where the operator $\mathcal{N}^{(k)}_{\mathrm{int}}$ is given by \eqref{oper Nint2}.
\end{theorem}

\begin{proof}
If $g\in\mathfrak{L}_{\gamma}(\mathcal{F}_\mathcal{H})$, mapping \eqref{krut_rozv} is defined
provided that $\gamma<e^{-1}$ and that the following estimate holds
\begin{eqnarray}\label{es}
  \big\|U^{+}(t)g\big\|_{\mathfrak{L}_{\gamma}(\mathcal{F}_\mathcal{H})}
  \leq e^2(1-\gamma e)^{-1}\| g \|_{\mathfrak{L}_{\gamma}(\mathcal{F}_\mathcal{H})}.
\end{eqnarray}
This estimate comes out from the inequality
\begin{eqnarray*}
  &&\big\|(U^{+}(t)g)_s\big\|_{\mathfrak{L}(\mathcal{H}_s)}\leq
  \sum_{n=0}^s\,\frac{1}{(s-n)!}\sum_{j_1\neq\ldots\neq j_{s-n}=1}^s \,
  \sum_{\mathrm{P}:\{Y\backslash X)_1,X\} =\bigcup_i X_i}(|\mathrm{P}|-1)!
  \|g_{s-n}\|_{\mathfrak{L}(\mathcal{H}_{s-n})}\leq\\
  &&\sum_{n=0}^s \|g_{s-n}\|_{\mathfrak{L}(\mathcal{H}_{s-n})}\frac{s!}{n!(s-n)!}
  \sum_{k=1}^{n+1}s(n+1,k)(k-1)!,
\end{eqnarray*}
where $s(n+1,k)$ are the Stirling numbers of the second kind for which it holds
\begin{eqnarray*}
        \sum\limits_{k=1}^{n+1}\texttt{s}(n+1,k)(k-1)!=\sum\limits_{k=1}^{n+1}\frac{1}{k}
        \sum\limits_{\substack{{r_{1},\ldots,r_{k}\geq 1}\\{r_{1}+\ldots +r_{k}=n+1}}}
        \frac{(n+1)!}{r_{1}!\ldots r_{k}!}\leq\sum\limits_{k=1}^{n+1}k^{n}
        \leq n!e^{n+2}.
\end{eqnarray*}

On the space $\mathfrak{L}_{\gamma}(\mathcal{F}_\mathcal{H})$
the $\ast$-weak continuity property  of the group $U^{+}(t)$ over the parameter $t\in \mathbb{R}^{1}$
is a consequence of the $\ast$-weak continuity of
group of operators \eqref{grG} for the Heisenberg equation \eqref{H-N1}.

In order to construct an infinitesimal generator of the group $\{U^{+}(t)\}_{t\in\mathbb{R}}$
we firstly differentiate the $nth$-term of expansion \eqref{krut_rozv}
in the sense of the pointweise convergence of the space $\mathfrak{L}_{\gamma}$.
If $g\in\mathcal{D}(\mathcal{N})\subset\mathfrak{L}_{\gamma}(\mathcal{F}_\mathcal{H})$, in a similar way to equality \eqref{Nint}
for $(1+n)th$-order dual cumulant \eqref{rozv_rec}, $n\geq1$, we derive
\begin{eqnarray}\label{derivation3}
       &&\lim\limits_{t\rightarrow 0}~\frac{1}{t}~\mathfrak{A}_{1+n}^{+} \big(t,(Y\backslash X)_1,X \big) g_{s-n}(Y\backslash X)\psi_{s}=\nonumber
       \sum\limits_{\mbox{\scriptsize $\begin{array}{c}{Z \subset Y\backslash X},\\{Z \neq\emptyset}\end{array}$}}
         \mathcal{N}_{\mathrm{int}}^{(|Z|+n)}(Z,X)g_{s-n}(Y\backslash X)\psi_{s}=\\
       &&\sum\limits_{k=1}^{s-n}\frac{1}{k!}\sum\limits_{i_1\neq\ldots\neq i_{k}\in\{j_1,\ldots,j_{s-n}\}}
         \mathcal{N}_{\mathrm{int}}^{(k+n)}(i_1,\ldots,i_{k},X)g_{s-n}(Y\backslash X)\psi_{s}.
\end{eqnarray}
Then, according to equalities \eqref{infOper1} and \eqref{derivation3}, for group \eqref{krut_rozv}  we obtain
\begin{eqnarray*}
   &&\lim\limits_{t\rightarrow 0}\frac{1}{t}\Big(\big(U^{+}(t)g\big) _{s}-g_{s}\Big)\psi_{s}
     = \lim\limits_{t\rightarrow 0}\frac{1}{t}\big(\mathfrak{A}_{1}^+(t)g_{s}-g_{s}\big)\psi_{s}+\\
   &&\sum_{n=1}^{s}\,\frac{1}{(s-n)!}\sum_{j_1\neq\ldots\neq j_{s-n}=1}^s\,\lim\limits_{t\rightarrow 0}\,
      \frac{1}{t}\,\,\mathfrak{A}_{1+n}^+ \big(t,(Y\backslash X)_1,X\big) g_{s-n}(Y\backslash X)\psi_{s}=\\
   &&\mathcal{N}_{s}g_{s}\psi_{s}+ \sum\limits_{n=1}^{s}\frac{1}{n!}
        \sum\limits_{k=n+1}^s \frac{1}{(k-n)!}\sum_{j_1\neq\ldots\neq j_{k}=1}^s
        \mathcal{N}_{\mathrm{int}}^{(k)}(j_1,\ldots,j_{k})g_{s-n}(Y\backslash\{j_1,\ldots,j_{n}\})\psi_{s},
\end{eqnarray*}
where we used identity \eqref{id}.

Thus if $g\in\mathcal{D}(\mathfrak{B}^{+})\subset\mathfrak{L}_{\gamma}(\mathcal{F}_\mathcal{H})$
in the sense of the $\ast$-weak convergence of the space $\mathfrak{L}_{\gamma}(\mathcal{F}_\mathcal{H})$ we finally have
\begin{eqnarray*}
\mathrm{w^{\ast}-}\lim\limits_{t\rightarrow 0}\big(\frac{1}{t}\big(U^{+}(t)g - g\big)
 -\mathfrak{B}^{+}g\big)=0,
\end{eqnarray*}
where the generator  ${\mathfrak{B}}^{+}={\bigoplus\limits}_{n=0}^{\infty}
{\mathfrak{B}}^{+}_{n}$ of group of operators \eqref{krut_rozv} is given by \eqref{d}.
\end{proof}

\subsection{The existence and uniqueness theorem}
The following theorem holds for abstract initial-value problem  \eqref{dh}-\eqref{dhi}
on the space $\mathfrak{L}_{\gamma}(\mathcal{F}_\mathcal{H})$.

\begin{theorem} A solution of the initial-value problem
to the quantum dual BBGKY hierarchy \eqref{dh}-\eqref{dhi} is determined by the expansion
\begin{equation}\label{sdh}
    G_{s}(t,Y)=\sum_{n=0}^s\,\frac{1}{(s-n)!}\sum_{j_1\neq\ldots\neq j_{s-n}=1}^s
      \mathfrak{A}_{1+n}^+ \big(t,(Y\backslash X)_1,X\big) \,
      G_{s-n}(0,Y\backslash X),  \quad\!\! s\geq 1,
\end{equation}
where the $(1+n)th$-order dual cumulant $\mathfrak{A}_{1+n}^+ \big(t,(Y\backslash X)_1,X\big)$
is defined by \eqref{rozv_rec}
\begin{equation*}
    \mathfrak{A}^{+}_{1+n}\big(t,(Y\backslash X)_1, X\big):=
    \sum\limits_{\mathrm{P}:\,\{(Y\backslash X)_1, X\}={\bigcup}_i X_i}
    (-1)^{\mathrm{|P|}-1}({\mathrm{|P|}-1})!\prod_{X_i\subset \mathrm{P}}\mathcal{G}_{|X_i|}(t,X_i),
\end{equation*}
where ${\sum}_\mathrm{P} $ is the sum over all possible partitions $\mathrm{P}$
of the set $\{(Y\backslash X)_1,j_1,\ldots,j_{s-n}\}$ into
$|\mathrm{P}|$ nonempty mutually disjoint subsets  $ X_i\subset \{(Y\backslash X)_1, X\}$.
For $G(0)\in\mathcal{D}({\mathfrak{B}}^{+})\subset\mathfrak{L}_{\gamma}(\mathcal{F}_\mathcal{H})$
it is a classical solution and for arbitrary initial data $G(0)\in\mathfrak{L}_{\gamma}(\mathcal{F}_\mathcal{H})$
it is a generalized solution.
\end{theorem}
\begin{proof} According to Theorem 1, for the initial data
$G(0)\in\mathcal{D}({\mathfrak{B}}^{+})\subset\mathfrak{L}_{\gamma}(\mathcal{F}_\mathcal{H})$, sequence \eqref{sdh}
is a classical solution of initial-value problem \eqref{dh}-\eqref{dhi} in the sense of the $\ast$-weak convergence of the space $\mathfrak{L}_{\gamma}(\mathcal{F}_\mathcal{H})$.

Let us now show that in the general case $G(0)\in\mathfrak{L}_{\gamma}(\mathcal{F}_\mathcal{H})$
expansions \eqref{sdh} give a generalized solution of
the initial-value problem to the quantum dual BBGKY hierarchy \eqref{dh}-\eqref{dhi}.
To this aim we consider the functional
\begin{eqnarray}\label{func-g}
\big(f,G(t)\big):= \sum_{s=0}^{\infty}\,\frac{1}{s!}
    \,\mathrm{Tr}_{\mathrm{1,\ldots,s}}\,f_{s}\,G_{s}(t),
\end{eqnarray}
where $f\in \mathfrak{L}^{1}_{\alpha,0}$  is a finite sequence of the degenerate trace class operators with infinitely times differentiable
kernels and with compact support. According to estimate \eqref{es} this functional exists
provided that $\alpha=\gamma^{-1}>e$ (see Section 4).

Using \eqref{sdh}, we can transform functional \eqref{func-g} as follows
\begin{eqnarray}\label{func-g-1}
\big(f,G(t)\big)=\big(f,U^{+}(t)G(0)\big)=\big(U(t)f,G(0)\big).
\end{eqnarray}
In this equality the group $U^{+}(t)$ is defined by expression \eqref{krut_rozv} (Theorem 1)
and $U(t)$ is an adjoint mapping to the group $U^{+}(t)$
\begin{eqnarray}\label{BBGKY-g}
(U(t)f) _{s}(Y)=\sum\limits_{n=0}^{\infty}\frac{1}{n!} \mathrm{Tr}_{\mathrm{s+1,\ldots,{s+n}}}
      \mathfrak{A}_{1+n}(t,Y_{1},X \backslash Y)f_{s+n}(X),
\end{eqnarray}
where $X\equiv\{1,\ldots,s+n\}$, i.e. $X \backslash Y\equiv\{s+1,\ldots,s+n\}$. In expansion \eqref{BBGKY-g}
the evolution operator $\mathfrak{A}_{1+n}(t,Y_{1},X \backslash Y)$ is
the $(1+n)th$-order cumulant of group of operators \eqref{groupG}
\begin{eqnarray*}\label{cum}
\mathfrak{A}_{1+n}(t,Y_{1},X \backslash Y)=
\sum\limits_{\mathrm{P}:\,\{Y_{1},X\setminus Y\}=
{\bigcup}_i X_i}(-1)^{|\mathrm{P}|-1}(|\mathrm{P}|-1)!
        \prod_{X_i\subset \mathrm{P}}\mathcal{G}_{|X_i|}(-t,X_i),
\end{eqnarray*}
where ${\sum}_\mathrm{P} $ is the sum over all possible partitions $\mathrm{P}$ of the set
$\{Y_{1},X\setminus Y\}=\{Y_{1},s+1,\ldots,s+n\}$ into $|\mathrm{P}|$ nonempty mutually disjoint subsets
$ X_i\subset \{Y_{1},X\setminus Y\}$.
If $f\in \mathfrak{L}^{1}_{\alpha,0}$, series \eqref{BBGKY-g} converges, provided that $\alpha > e$ \cite{GerS}
and the functional $\big(U(t)f,G(0)\big)$ exists.

The one-parameter family of operators $U(t)$ is differentiable with respect to $t$ and for $f\in\mathfrak{L}^{1}_{\alpha,0}$
an infinitesimal generator of group \eqref{BBGKY-g} is defined by the following expression (see also \eqref{1b})
\begin{equation} \label{1}
\begin{split}
        &(\mathfrak{B}f)_{s}(Y)=
        -\mathcal{N}_{s}(Y)f_{s}(Y)+\\
        &\sum\limits_{k=1}^{s}\frac{1}{k!}\sum\limits_{i_1\neq\ldots\neq i_{k}=1}^{s}
        \,\sum\limits_{n=1}^{\infty}\frac{1}{n!}\mathrm{Tr}_{\mathrm{s+1,\ldots,s+n}}
    \big(-\mathcal{N}_{\mathrm{int}}^{(k+n)}\big)(i_1,\ldots, i_{k},X \backslash Y)f_{s+n}(X).
\end{split}
\end{equation}

Since for bounded interaction potentials \eqref{H_Zag}, if $f\in \mathfrak{L}^{1}_{\alpha,0}$,
the operator $\mathfrak{B}U(t)f$ is a trace class operator, the
operator $\mathfrak{B}U(t)f G(0)$ is also a trace class operator then the functional $\big(\mathfrak{B}U(t)f, G(0)\big)$
exists. Moreover, it holds the equality: $\big(\mathfrak{B}U(t)f,G(0)\big)=\big(U(t)\mathfrak{B}f,G(0)\big)$ and the following
result
\begin{equation*}
\begin{split}
&\lim\limits_{t\rightarrow 0}\big|\big(\frac{1}{t}(U(t)-I)f,G(0)\big)-\big(\mathfrak{B}f,G(0)\big)\big|=\\
&\lim\limits_{t\rightarrow 0}\big|\sum_{s=0}^{\infty}\,\frac{1}{s!}
    \,\mathrm{Tr}_{\mathrm{1,\ldots,s}}\big(\frac{1}{t}\big(U(t)f-f\big)_{s}G_{s}(0)-\big(\mathfrak{B}f\big)_{s}G_{s}(0)\big)\big|\leq\\
&\big\|G(0)\big\|_{\mathfrak{L}_{\gamma}(\mathcal{F}_\mathcal{H})}\,\lim\limits_{t\rightarrow 0}\, \sum_{s=0}^{\infty}\,{\gamma}^{-s}\,\mathrm{Tr}_{\mathrm{1,\ldots,s}}
    \big|\frac{1}{t}\big(U(t)f-f\big)_{s}-(\mathfrak{B}f)_{s}\big|=\\
    &\big\|G(0)\big\|_{\mathfrak{L}_{\gamma}(\mathcal{F}_\mathcal{H})}\,\lim\limits_{t\rightarrow 0}\big\|\frac{1}{t}\big(U(t)f-f\big)-\mathfrak{B}f \big\|_{\mathfrak{L}^{1}_{{\gamma}^{-1}}(\mathcal{F}_\mathcal{H})}=0.
\end{split}
\end{equation*}
Hence equality \eqref{func-g-1} can be differentiated with respect to time and we get finally
\begin{equation*}
\begin{split}
&\frac{d}{dt}\big(f,G(t)\big)=
\big(U(t)\mathfrak{B}f,G(0)\big)=
\big(\mathfrak{B}f,U^{+}(t)G(0)\big)=\\
&\big(\mathfrak{B}f,G(t)\big),
\end{split}
\end{equation*}
where the operator $\mathfrak{B}$ is defined by \eqref{1}.

These equalities mean that the sequence of operators \eqref{sdh} for arbitrary $G(0)\in\mathfrak{L}_{\gamma}(\mathcal{F}_\mathcal{H})$
is a generalized solution of the Cauchy problem to the quantum dual BBGKY hierarchy \eqref{dh}-\eqref{dhi}.
\end{proof}

\section{The existence of the mean value observable\\ functional}
As it was above mentioned, the functional of the mean value \eqref{avmar}
defines a duality between marginal observables and marginal states.
If $G(t)\in\mathfrak{L}_{\gamma}(\mathcal{F}_\mathcal{H})$ and $F(0)\in \mathfrak{L}^{1}_{\alpha}$ then,
according to estimate \eqref{es}, the functional
\begin{equation}\label{avmar-1}
       \big\langle G(t)\big|F(0)\big\rangle=
        \sum\limits_{s=0}^{\infty}\,\frac{1}{s!}\,
        \mathrm{Tr}_{\mathrm{1,\ldots,s}}\,G_{s}(t,1,\ldots,s)F_{s}(0,1,\ldots,s).
\end{equation}
exists, provided that $\alpha=\gamma^{-1}>e$, and the following estimate holds
\begin{equation*}
      \big|\big\langle G(t)\big|F(0)\big\rangle\big|\leq e^2(1-\gamma e)^{-1}\big\|G(0)\big\|_{\mathfrak{L}_{\gamma}(\mathcal{F}_\mathcal{H})}
      \big\|F(0)\big\|_{\mathfrak{L}^{1}_{{\gamma}^{-1}}(\mathcal{F}_\mathcal{H})}.
\end{equation*}
Thus, marginal density operators from the space $\mathfrak{L}^{1}_{\alpha}$ describe finitely many quantum particles.
Indeed, for such additive-type observable as number of particles, i.e. one-component sequence $N(0)=\big(0,I,0,\ldots\big)$,
according to definition \eqref{cumulant} of dual cumulants, the expansion for solution \eqref{hsol} get the following form (see also \eqref{af})
\begin{eqnarray*}
    \big(N(t)\big)_{s}(Y)= \mathfrak{A}_{s}^{+}(t,1,\ldots,s)
       \sum_{j=1}^s I= I \delta_{s,1}, \quad s\geq 1
\end{eqnarray*}
and we have
\begin{equation*}
      \big|\big\langle N(t)\big|F(0)\big\rangle\big|=\big|\mathrm{Tr}_{\mathrm{1}}\,F_{1}(0,1)\big|\leq \big\|F(0)\big\|_{\mathfrak{L}^{1}_{{\gamma}^{-1}}(\mathcal{F}_\mathcal{H})}<\infty.
\end{equation*}

We have above stated the properties of group \eqref{krut_rozv} defined on $\mathfrak{L}_{\gamma}(\mathcal{F}_\mathcal{H})$.
In order to describe the evolution of infinitely many particles \cite{CGP97}
the problem lies in the definition of functional \eqref{avmar-1} for operators from the suitable Banach spaces. Namely,
marginal density operators have to belong to more general spaces than  $\mathfrak{L}^{1}_{\alpha}(\mathcal{F}_\mathcal{H})$.
For example, it can be the space of sequences of bounded operators containing the equilibrium states \cite{Gen}.
In this case every term of expansions for the mean value functional \eqref{avmar-1}
contains the divergent traces \cite{CGP97},\cite{BG},\cite{GerS} and the analysis of such a question
for quantum systems remains an open problem.

\section{Conclusion}

The concept of dual cumulants \eqref{cumulant} of groups of operators \eqref{grG} for the Heisenberg equations \eqref{H-N1}
or cumulants \eqref{BBGKY-g} of groups of operators \eqref{groupG} for the von Neumann equations \eqref{F-N1}
forms the basis of the groups of operators for quantum evolution equations as well as the quantum dual BBGKY hierarchy and
the BBGKY hierarchy for marginal density operators \cite{GerS}.

In the case of quantum systems of particles,
obeying Fermi or Bose statistics, group of operators \eqref{krut_rozv} has different structures.
The analysis of these cases will be given in a separate paper.

On the space $\mathfrak{L}_{\gamma}(\mathcal{F}_\mathcal{H})$ one-parameter mapping \eqref{krut_rozv} is not a strong continuous group.
The group $\{U^{+}(t)\}_{t\in\mathbb{R}}$ of operators \eqref{krut_rozv} defined on the space $\mathfrak{L}_{\gamma}(\mathcal{F}_\mathcal{H})$
is dual to the strong continuous group $\{U(t)\}_{t\in\mathbb{R}}$ of operators \eqref{BBGKY-g} for the BBGKY hierarchy defined on
the space $\mathfrak{L}_{\alpha}^{1}(\mathcal{F} _\mathcal{H})$ and the fact that it is a $C_{0}^{\ast}$-group
follows also from general theorems about properties of the dual semigroups \cite{BR , Pazy}.

We have constructed infinitesimal generator \eqref{d} on the domain
$\mathcal{D}(\mathfrak{B^+})\subset\mathfrak{L}_{\gamma}(\mathcal{F}_\mathcal{H})$ which is everywhere
dense set for the $\ast$-weak topology of the space $\mathfrak{L}_{\gamma}(\mathcal{F}_\mathcal{H})$.
The question of how to define the domain of the definition $\mathcal{D}(\mathfrak{B^+})$
of generator \eqref{d} is an open problem \cite{BR, AL}.

\section*{Acknowledgement}
This work was performed under the auspices of the {\it National
Group for Mathematical Physics} of the {\it Istituto Nazionale di
Alta Matematica} and was partially supported by the WTZ grant No M/124 (UA 04/2007), by the {\it Italian
Ministery of University} MIUR National Project {"Kinetic and Hydrodynamic
Equations of Complex Collisional Systems", PRIN 2006)} and
by Research Funds of the University of Florence (Italy).

\end{document}